\documentclass[conference]{IEEEtran}
\IEEEoverridecommandlockouts
\pdfoutput=1
\usepackage{cite}
\usepackage{amsmath,amssymb,amsfonts}
\usepackage[ruled, vlined,linesnumbered]{algorithm2e}
\usepackage{mathtools}
\usepackage{graphicx}
\usepackage{textcomp}
\usepackage{xcolor}
\usepackage{hyperref}
\usepackage{cleveref}
\usepackage{float}

\usepackage{tikz}
\usetikzlibrary{positioning}
\usetikzlibrary{arrows.meta}
\usetikzlibrary{calc}
\usetikzlibrary{shapes.misc}
\usetikzlibrary{decorations.pathreplacing}
\usetikzlibrary{matrix}
\usetikzlibrary{shapes.geometric}

\usepackage[USenglish]{babel}
\usepackage[babel]{csquotes}
\usepackage{wrapfig}
\usepackage{booktabs}
\usepackage[per-mode=symbol,binary-units=true]{siunitx}[=v2]
\DeclareSIUnit{\nothing}{\relax}
\usepackage{collcell}
\usepackage{multirow}

\DeclareMathOperator{\rank}{rank}
\DeclareMathOperator{\BigO}{\mathcal{O}}

\newif\ifmicrotype
\newif\ifpagenumbers
\pagenumbersfalse

\microtypetrue

\ifmicrotype
\usepackage[%
activate={true,nocompatibility},%
final,%
tracking=true,%
kerning=true,%
spacing=true,%
stretch=10,%
shrink=30]{microtype}
\microtypecontext{spacing=nonfrench}
\SetTracking{encoding={*}, shape=sc}{0}
\fi

\makeatletter
\def\ps@IEEEtitlepagestyle{%
  \def\@oddfoot{\mycopyrightnotice}%
  \def\@evenfoot{}%
}
\def\mycopyrightnotice{%
    {\footnotesize 
        \begin{minipage}{\textwidth}
        \copyright~2023 IEEE. Personal use of this material is permitted. Permission
from IEEE must be obtained for all other uses, in any current or future
media, including reprinting/republishing this material for advertising or
promotional purposes, creating new collective works, for resale or
redistribution to servers or lists, or reuse of any copyrighted
component of this work in other works. Published version: \href{https://doi.org/10.1109/IPDPS54959.2023.00076}{10.1109/IPDPS54959.2023.00076}~\cite{Sanders2023}.
        \end{minipage}
}
  \gdef\mycopyrightnotice{}
}
\makeatother

\begin{document}

\nocite{Sanders2023}

\title{
	Engineering a Distributed-Memory \\ Triangle Counting Algorithm
\thanks{
    \begin{wrapfigure}{R}{.33\columnwidth}
      \vspace{-1.25\baselineskip}
      \includegraphics[width=.33\columnwidth]{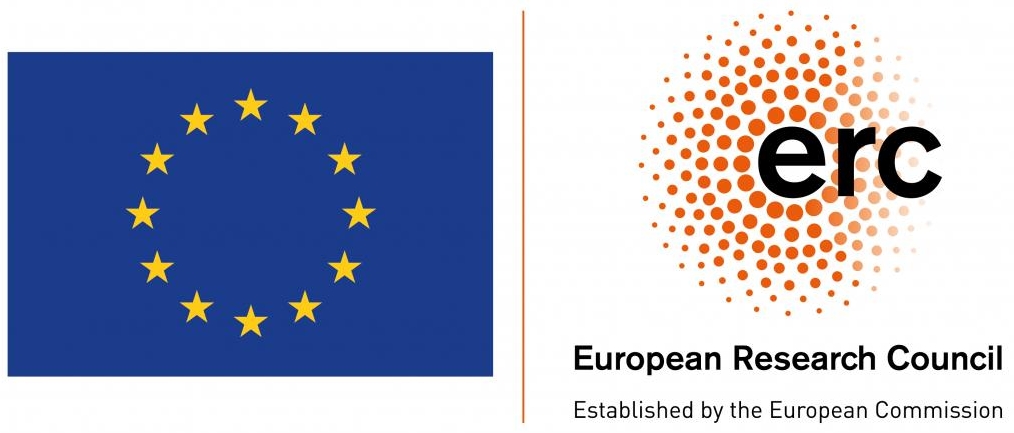}
    \end{wrapfigure}	

    \noindent This project has received funding from the European Research
    Council (ERC) under the European Union’s Horizon 2020 research and
    innovation programme (grant agreement No. 882500).

}
}

\author{
\IEEEauthorblockN{Peter Sanders}
\IEEEauthorblockA{
  \textit{Institute of Theoretical Informatics}\\
	\textit{Karlsruhe Institute of Technology}\\
  Karlsruhe, Germany\\
	sanders@kit.edu
}
\and
\IEEEauthorblockN{Tim Niklas Uhl}
\IEEEauthorblockA{
  \textit{Institute of Theoretical Informatics}\\
	\textit{Karlsruhe Institute of Technology}\\
  Karlsruhe, Germany\\
	uhl@kit.edu
}
}

\maketitle
\ifpagenumbers
\pagestyle{plain}
\fi

\DontPrintSemicolon

\makeatletter
\newcommand{\removelatexerror}{\let\@latex@error\@gobble}
\newcommand{\frage}[1]{{\sf[#1]}}
\renewcommand{\ps}[1]{{\frage{\color{red}ps: #1}}}
\renewcommand{\nu}[1]{{\frage{\color{blue}nu: #1}}}

\makeatother

\newtheorem{Lemma}{Lemma}
\newtheorem{corollary}{Corollary}

\DontPrintSemicolon
\SetKwData{Sum}{SUM}
\SetKwFunction{Red}{Reduce}
\SetKw{Send}{send}
\SetKw{To}{to}
\SetKw{Recv}{receive}
\SetKwFunction{AllToAll}{SparseAllToAll}
\SetKwFor{OnReceive}{on receive}{do}{end}
\SetKwProg{Fn}{function}{}{end}
\SetKwFunction{CountTriangles}{count\_triangles}
\SetKwFunction{ExDeg}{exchange\_ghost\_degree}
\SetKwFunction{Emit}{emit\_triangle}

\begin{abstract}
  Counting triangles in a graph and incident to each vertex is a
  fundamental and frequently considered task of graph analysis.  We
  consider how to efficiently do this for huge graphs using massively
  parallel distributed-memory machines. Unsurprisingly, the main
  issue is to reduce communication between processors. We achieve this
  by counting locally whenever possible and reducing the amount of
  information that needs to be sent in order to handle (possible)
  nonlocal triangles.  We also achieve linear memory requirements
  despite superlinear communication volume by introducing a new
  asynchronous sparse-all-to-all operation. Furthermore, we
  dramatically reduce startup overheads by allowing this communication
  to use indirect routing.  Our algorithms scale (at least) up to 32\,768 cores 
  and are up to 18 times faster than the previous
  state of the art.
\end{abstract}


\section{Introduction}
\label{sec:intro}

Graphs are a universally used abstraction to describe relations
between objects and thus a key to many applications of computers. With
the big-data aspect of the information age, it is therefore clear that
processing very large graphs quickly and efficiently is an
increasingly important problem. For example, the largest social
network, Facebook, had over 2.7 billion active users in the fourth
quarter of 2020~\cite{Statista2021}, and the Common Crawl 2012 web
hyperlink graph~\cite{Meusel} consists of over 100 billion
edges and requires 350GB of memory in uncompressed form. A description
of the neural connections of the human brain would have about 10
billion vertices (neurons) and 100 trillion edges (synapses). The need
to use supercomputers for processing such large networks is widely
recognized as witnessed by the popularity of the Graph 500
benchmark~\cite{Graph}. However, the number of graph problems that
can be handled efficiently with many thousands of processing elements
(PEs) is quite limited so far. Graph 500 concentrates mostly on
breadth-first search (BFS) on a single kind of input that is very easy
for BFS.\footnote{The static graph challenge~\cite{Samsi2017} includes a triangle counting benchmark
but most entries so far do not look at the largest available machines.
Similarly, graph processing tools frequently benchmark triangle counting
but currently do not scale very well; see also Section~\ref{sec:related}.}

In this paper we want to improve this situation for one of the most
widely used graph-analysis problems -- \emph{triangle counting}. Given an
undirected graph $G=(V,E)$, we are looking for the number of sets
$\{u,v,w\}\subseteq V$ such that these three vertices are mutually
connected in $E$. Since triangles can be very non-uniformly distributed
over that graph, one often also wants to know the number of triangles incident
to each vertex -- normalized to the range $[0,1]$. This is known as the
\emph{local clustering coefficient (LCC)}.

Triangles
are the smallest non-trivial complete subgraph and often indicative of dense
regions of the graph. Therefore, the problem has numerous applications
in analyzing complex networks such as social
graphs~\cite{Newman2003} but is also applicable in
practice. Becchetti et al.~\cite{Becchetti2008} show that
analyzing the distribution of the local clustering coefficient may be
used to detect spam pages on the web, and Eckmann and
Moses~\cite{Eckmann2002} identify common topics of web pages using
triangle counting. Other applications include database query
optimizations~\cite{BarYossef2002} and link
recommendation~\cite{Tsourakakis2011}.

After introducing basic concepts in Section~\ref{sec:prelim}, we discuss previous approaches in Section~\ref{sec:related}. Our algorithms are described in Section~\ref{sec:algorithms} and evaluated in Section~\ref{sec:evaluation}. Section~\ref{sec:conclusion} summarizes the results and outlines possible future work.


\paragraph*{Contributions}
\begin{itemize}
\item All triangles that can be found locally are found locally.
\item Only cut edges need to be communicated.
\item Linear memory requirements using an asynchronous sparse all-to-all algorithm.
\item Fast and highly scalable MPI code.
\item Better scalability by indirect communication.
\item Extensive experiments with large real-world inputs and
  massive generated graphs from several families of inputs.
\item Comparison with state-of-the art competitors shows
   up to 18$\times$ better performance for
  large configurations.
\item Generalization to exact and approximate computation of local
  clustering coefficients.
\end{itemize}

\section{Preliminaries}
\label{sec:prelim}
In this section we introduce the basic concepts and notations used in this work. We further provide details of the underlying machine model.

\subsection{Basic Definitions}
\label{subsec:definitions}
We consider undirected graphs $G=(V, E)$  where $V = \{0, \ldots, n - 1\}$ 
and $E \subseteq \binom{V}{2}$ are the sets of vertices and edges, respectively. 
G has $n = |V|$ vertices and $m = |E|$ edges. 
For a given vertex $v \in V$ let $N_v(G) = \{u \in V \mid \{u, v\} \in E\}$ denote the \emph{neighborhood} of $v$. 
The \emph{degree} of $v$ is $d_v \coloneqq |N_v(G)|$. If the considered graph $G$ is clear from the context, we simply write $N_v$ instead of $N_v(G)$.

Two vertices $u, v \in V$ are called \emph{adjacent} if the edge $\{u, v\} \in E$ exists. 
An edge $e \in E$ is \emph{incident} to a vertex $v \in V$ if $v \in e$. 
The vertices incident to an edge are called \emph{endpoints}.

For $V' \subseteq V$ we define the \emph{induced subgraph} $G(V') = (V', E')$ of $G$, where $E' \coloneqq \{ \{u, v\} \in E \mid u \in V' \land v \in V'\}$.
For a set of three distinct vertices $u, v, w \in V$, we call the induced subgraph $G(\{u, v, w\})$ a \emph{triangle} if and only if it is complete, i.e., each edge $\{u, v\}, \{v, w\}, \{w, u\}$ exists.

Most triangle counting algorithms depend on orienting undirected graphs, to prevent redundant counting of triangles. This is accomplished by using a total ordering $\prec$ on the vertices. Each edge is directed from the lower to higher ranked vertex.
With respect to a total ordering $\prec$, we define the outgoing neighborhood of $v \in V$ as $N_v^+(G) \coloneqq \{u \in V \mid \{v, u\} \in E \land v \prec u\}$ and the incoming neighborhood as $N_v^-(G) \coloneqq N_v(G) \setminus N_v^+(G)$. We write a directed edge as an (ordered) tuple $(u, v)$ if $u \prec v$. We define the out-degree $d_v^+$ of a vertex $v \in V$ as $d_v^+ \coloneqq d_v^+(G) \coloneqq |N_v^+(G)|$. As before, we do not mention the considered graph explicitly if it is clear from the context. We call each pair of directed edges $(v, u)$, $(v, w)$ a \emph{wedge}. A naive triangle counting algorithm would be to enumerate all wedges and check if a closing edge $(u, w)$ or $(w, u)$ exists.

\subsection{Machine Model and Input Format}
\label{sec:machine-model}

We consider a system consisting of $p$ processing elements (PEs) numbered $P_0, \ldots, P_{p - 1}$, which are connected via a network with full-duplex, single-ported communication.
Sending a message of length $\ell$ from one PE to another takes time $\alpha + \beta\ell$, where $\alpha$ is the time required to initiate a connection and $\beta$ the subsequent transmission time for sending one machine word. Let the \emph{communication volume} denote the total number of machine words sent between processors.

We assume that input graphs are stored in the \emph{adjacency array} format, which stores the set of neighbors $N_v$ for each vertex $v$ using two arrays in a compressed form.
We use \emph{1D partitioning}, which means that each $P_i$ is assigned a vertex set $V_i$, where all sets $V_i$ are disjoint and $V = \bigcup_{i=0}^{p - 1} V_i$.
$P_i$ stores the neighborhoods for a subsequence $V_i$ of the vertices $\{0, \ldots, n - 1\}$. A vertex is called \emph{local} to $P_i$ if it lies in $V_i$. For a vertex $v$ the rank is defined as $\rank(v) = i :\Leftrightarrow v \in V_i$.
We assume that processor $P_i$ only has access to vertices in $V_i$ and their neighborhoods.

In addition to that we assume that the vertices are globally ordered among the processors by vertex ID. This means that if $\text{rank}(v) < \text{rank}(w)$ for vertices $v \in V_i$, $w \in V_j$, $i \neq j$ then $w$ has a higher ID than $v$.

Vertices $u$ which are contained in a neighborhood $N_v$ for a vertex $v \in V_i$, but are not local to $P_i$ are called \emph{ghost vertices}. Local vertices which are adjacent to at least one ghost vertex are called \emph{interface vertices}.
An edge connecting two vertices $v \in V_i$, $w \in V_j$, $i \neq j$ is called \emph{cut edge}.
We define the \emph{cut graph} $\partial G$ as the graph only consisting of cut edges, i.e., $\partial G \coloneqq (V, E')$, where $E' \coloneqq \{{u, v} \in E \mid u \in V_i, v \in V_j, i \neq j\}$.
For processor $P_i$ let $\overline{V_i} \coloneqq V_i \cup \bigcup_{v \in V_i} N_v$ denote the set of all local vertices and ghost vertices and let $\partial V_i \coloneqq \overline{V_i} \setminus V_i$ denote the set of ghost vertices. 

\begin{figure}
	\centering
	\includegraphics[page=5]{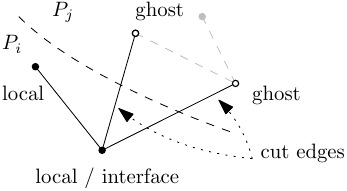}
	\caption{Example of the various vertex types for the local graph view from $P_i$. The edges and vertices colored in gray are not visible to $P_i$.}
	\label{fig:vertex-types}
\end{figure}

An example of the terminology used is given in Fig.~\ref{fig:vertex-types}.

\section{Related Work}\label{sec:related}
With growing size of real world input instances, there was a correspondingly rising need for efficient triangle counting algorithms.
A large variety of approaches has been proposed, tailored to different models of computation.
As for sequential algorithms, 
Schank's Ph.D. thesis~\cite{Schank2007} and 
a more recent work by Ortmann and Brandes~\cite{Ortmann2014} give an extensive overview.

Almost all triangle counting algorithms derive from a single sequential base algorithm, which is often referred to as \textsc{EdgeIterator}.
It iterates over all edges $\{u, v\}$ in the graph and intersects the neighborhoods of both endpoints.
This counts each triangle three times, once from each contained edge.
A practical formulation called \textsc{compact-forward} which avoids redundant counting is attributed to Latapy~\cite{Latapy2008}. The algorithm orients the undirected input graph using a degree-based (total) ordering defined as follows: For $u, v \in V$
\[
u \prec v \Leftrightarrow 
\begin{cases}
	d_u < d_v 	& \text{if } d_u \neq d_v \\
	u < v 		& \text{if } d_u = d_v\text{\;.}
\end{cases}
\]
By only considering outgoing neighborhoods, this avoids finding duplicate triangles and 
also reduces the overall work, because the out-degree of high-degree vertices is reduced.
Pseudocode is given in Algorithm~\ref{alg:edge-iterator}.

\begin{figure}[b]
{
	\removelatexerror
	\begin{algorithm}[H]
		
		$T \leftarrow 0$\;
		\For{$v \in V$}{%
			\For{$u \in N_v^+$}{%
				$T \leftarrow T + |N_v^+ \cap N_u^+|$\; \label{line:intersection-edge-iterator}
			}
		}
		\caption{\textsc{EdgeIterator}}
		\label{alg:edge-iterator}
	\end{algorithm}
}
\end{figure}

The set intersection relies on the neighborhoods being sorted.
It is implemented using a procedure similar to the merge phase of merge sort.

\subsection{Parallel Algorithms}
Even carefully tuned sequential algorithms are not enough for today's massive problem instances. To deal with large inputs, one has to exploit processor and memory parallelism.
\subsubsection{Shared Memory}
On a shared memory machine, \textsc{EdgeIterator} may be easily parallelized. As all set intersections in line~\ref{line:intersection-edge-iterator} are independent of each other they may be carried out in parallel and lock-free, as each thread has random access to the whole graph and does not modify it. Shun and Tangwongsan~\cite{Shun2015} propose a parallel version of \textsc{EdgeIterator}, where the loops in lines~\ref{line:loop_vert}-\ref{line:loop_n} over vertices $v \in V$ and vertices $u \in N_v^+$ are executed in parallel. Dhulipala, Shun and Blelloch~\cite{Dhulipala2021} extend this approach to work on large compressed graphs. They  also parallelize the set intersection. 

Other shared memory implementations~\cite{Tom2017, Rahman2013} share the same main idea with Shun and Tangwongsan's algorithm: They are based on variants of \textsc{EdgeIterator} and execute the two outer loops of the algorithms in parallel. 

Green et al.~\cite{Green2014} use a different approach. Instead of parallelizing on node level, they iterate over all local edges in parallel and intersect the neighborhoods of the endpoints. They estimate the work required per edge and statically partition the edge list into chunks of equal work. They report that if not taking the partitioning step into account, this approach outperforms the node-centric parallelization strategy.

\subsubsection{Distributed Memory}
\label{sec:distributed-memory}
While triangle counting is easily parallelized on shared memory machines, the availability of systems with high number of processors and sufficient amount of memory for large graphs is still limited today. 
Distributed memory machines provide us with a large total amount of memory and abundant parallelism.
Under the assumption that the graph is 1D partitioned as described in Section~\ref{sec:machine-model}, \textsc{EdgeIterator} can be adapted to this distributed setting as shown in Algorithm~\ref{alg:dist-edge-iterator}. When processing a local edge $(v, u)$ the local neighborhoods are intersected as before, but when a cut edge is encountered, the neighborhood $N_v^+$ of $v$ is sent to the local rank $P_j$ of $u$, which then performs the set intersection upon receiving the message. The total number of triangles in the graph is then obtained by reducing over the local triangle counts. 
Firstly, consider that $P_i$ processes two edges $(v, u), (v, u^\prime)$ 
with $\rank(u) = \rank(u^\prime) \neq \rank(v)$. 
Then $N_v^+$ would be sent to PE $\rank(u)$ twice. 
Arifuzzaman et. al~\cite{Arifuzzaman2015} address this issue 
and ensure that each neighborhood is only sent to each PE once 
by exploiting the sortedness of vertex neighborhoods 
and require only $\BigO(|V_i|)$ additional memory per PE.

\begin{figure}[t]
	\removelatexerror
	\begin{algorithm}[H]
		$T_i \leftarrow 0$\;
		\For{$v \in V_i$}{\label{line:loop_vert}%
			\For{$u \in N_v^+$}{\label{line:loop_n}%
				\lIf{$u \in V_i$} {
					$T_i \leftarrow T _i+ |N_v^+ \cap N_u^+|$
				} \lElse {
					\Send $((v, u), N_v^+)$ \To $P_{\text{rank}(u)}$
				}
				
			}
			
		}
		\OnReceive{$((v, u), N_v^+)$}{
			$T_i \leftarrow T _i+ |N_v^+ \cap N_u^+|$\;
		}
		$T \leftarrow \Red{$T_i$, \Sum}$\;
		\caption{Distributed-memory \textsc{EdgeIterator}}
		\label{alg:dist-edge-iterator}
	\end{algorithm}
\end{figure}

Still, their approach sends many small messages, which may lead to high startup overheads. Ghosh et al.~\cite{Ghosh2020} address this by aggregating messages and then finally perform a single all-to-all collective operation.
Because they only use a single communication step and never empty the buffer, it may require memory superlinear in the input size.

There also exist distributed algorithms based on matrix multiplication on the adjacency matrix, but it is shown that they only scale up to a couple of hundred PEs~\cite{Tom2019, Azad2015}.

Suri and Vassilvitskii~\cite{Suri2011} describe two distributed algorithms using the popular \emph{MapReduce} framework~\cite{Dean2004} which are both based on \textsc{EdgeIterator}. Park et al.~\cite{Park2014} further improve this algorithm, but all algorithms produce large amounts of intermediate data by replicating the input graph and in turn require a lot of communication during the shuffle phase, which hinders scalability~\cite{Afrati2012, Park2014}.

Pearce et al.~\cite{Pearce2014, Pearce2017, Pearce2019} present a triangle counting algorithm using their distributed vertex-centric framework HavoqGT. On the degree-oriented graph they generate all open wedges $(u, v, w)$ for each vertex $v \in V$, i.e., all pairs of outgoing neighbors $\{u, w\} \in \binom{N_v^+}{2}$, and create new vertex visitor for these neighbors, which then check for a closing edge $(u, w)$ or $(w, u)$. They partition the neighborhoods of  high-degree vertices among multiple PEs and also employ message aggregation. To reduce the number of messages, they first aggregate messages at node level and then reroute them to other compute nodes.

There exist algorithms which avoid communication entirely during the counting
phase by replicating complete neighborhoods of ghost vertices during
preprocessing~\cite{Hoang2019, Arifuzzaman2013}. This basically offloads the
communication done by other approaches to the preprocessing phase and requires
superlinear memory to store the replicated vertex neighborhoods. This limits the
overall scalability for large inputs~\cite{Arifuzzaman2015}.

\subsection{Approximative Algorithms}
\label{subsec:approx}
For many applications it suffices to only approximate the number of triangles instead of determining the exact result. 
Approximation algorithms may reduce both the time and memory requirements of triangle counting when an exact result is not required.
Tsourakakis et al.~\cite{Tsourakakis2009} introduce \textsc{Doulion}, an
edge-sampling-based approximation algorithm that reduces the input size. Instead
of sampling edges independently, Pagh and Tsourakakis~\cite{Pagh2012} achieve better approximations by coloring vertices independently and only considering the graph of edges where both endpoints have the same color.
Both approaches require a (distributed) triangle counting algorithm as a black box to count the triangles in the reduced graph and scale the result accordingly to obtain an approximation.
There also exist semi-streaming algorithms for approximating triangle counts~\cite{BarYossef2002, Jha2013}.

\subsection{Miscellaneous}
For a broader overview on various other specialized algorithms, we refer the reader to the survey by Al Hasan and Dave~\cite{AlHasan2018}. While we focus on message passing based approaches for distributed-memory machines there has also been a lot of work using other models of computation.
GPU algorithms~\cite{Green2014a, Hu2018, Hoang2019, Pandey2019} focus on parallelizing the set intersection operation using binary search based approaches, which are more suitable to be implemented on GPUs than merge-based approaches. 
The external memory algorithm by Chu and Cheng~\cite{Chu2011} uses a locality aware contraction technique which is similar to ours. 
They use it to load chunks of the input graph from disk, count local triangles, remove internal edges and write the contracted graph chunk back to disk.
This is repeated until the whole contracted graph fits into main memory. 
The advances in GPU and external settings are orthogonal to our approach, because they may be used to count local triangles if each PE is equipped with a GPU or the main memory is limited. Our proposed communication reduction techniques may still be employed.

\section{Our Algorithms}\label{sec:algorithms}
We identify reducing communication as one of the key challenges for designing a scalable distributed memory triangle counting algorithm.
Recall that sending a single message of length $\ell$ takes time $\alpha +
\beta\ell$ in the full-duplex model of communication.
 
We can therefore reduce communication by addressing two parts of this communication model:
By limiting the number of messages and reducing the total startup overhead or by reducing the total communication volume.
In this section we present two algorithms which address both. 

\textsc{Ditric} (\textbf{Di}stributed \textbf{Tri}angle \textbf{C}ounting)
employs message aggregation with linear memory requirements and introduces a
network agnostic indirect communication protocol.

\textsc{Cetric} (the \textbf{c}ommunication-\textbf{e}fficient variant of \textsc{Ditric}) builds upon this, exploits locality and
uses graph contraction such that the communication volume is only dependent on
the structure of the cut graph. In section~\ref{sec:extensions} we show how we
can reduce communication even more when an approximation of the triangle count
is acceptable.

\subsection{Message Aggregation}
\begin{figure}
	\centering
	\input{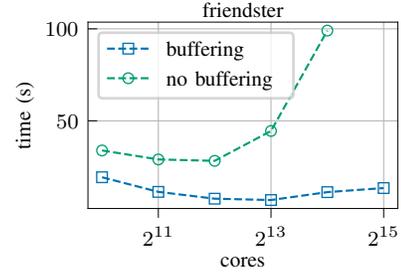}
	\vspace*{-10pt}
	\caption{Running time of the basic distributed algorithm on friendster with and without message aggregation.}
	\label{fig:buffering}
\end{figure}
As already mentioned in Section~\ref{sec:distributed-memory}, a direct adaptation of \textsc{EdgeIterator} 
to a distributed system leads to a high number of messages between PEs.
To reduce the startup overhead required for sending a message it is feasible to 
aggregate multiple small messages designated for the same receiver into a single one.

An example how message aggregation improves on scalability is shown in Fig.~\ref{fig:buffering}, 
where we compare the running time of distributed \textsc{EdgeIterator} with and
without message aggregation enabled on the friendster graph.
 
A challenge for distributed triangle counting is that 
the total communication volume is superlinear in the input size,
because each neighborhood may be sent to multiple other PEs.
This is a particular issue when using message aggregation since local buffers
can already overflow when just one PE needs to send too many messages.

Our algorithm \textsc{Ditric} uses a dynamically buffered message queue to solve this problem. Let $\delta$ be the threshold upon which the buffer should be emptied.
Each PE maintains a hash-map of buffers $B_j$ as a dynamic array for each communication partner $P_j$. 
If a send operation of the neighborhood of a vertex $v$ to PE $P_j$ is issued, we append the neighborhood to the buffer $B_j$.
If this operation results in the overall buffer size $B = \sum_{j = 0}^{p-1} |B_j|$ to become greater than the threshold $\delta$, we send all $B_j$ to their corresponding receiver PEs. We do this by using double buffering: We replace each $B_j$ with an empty buffer and pass the full buffer to the MPI runtime by issuing a non-blocking send operation. While the send operation is carried out, we can continue writing messages to the now empty buffer $B_j$ and only block in the unlikely event that the second buffer overflows while the first buffer has not been completely sent yet.

Each PE continuously polls for incoming messages and processes them.

By setting $\delta \in \BigO(|E_i|)$ we ensure that the memory required per PE never exceeds the local input size.

\subsection{Indirect Message Delivery}
\label{sec:indirect}
\begin{figure}[b]
	\centering
	\includegraphics[page=6, width=.5\columnwidth]{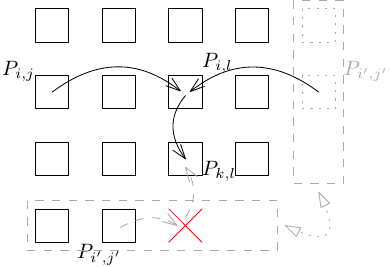}
	\caption{Grid-based indirect message delivery in \textsc{Cetric}.}
	\label{fig:indirect}
\end{figure}
Message aggregation helps to reduce the startup overhead when a PE sends many small neighborhoods to another communication partner. 
If a PE owns a high-degree vertex, it  has to send and receive many small vertex neighborhoods from/to different communication partners. 
We propose a simple grid-based indirection scheme, which, combined with
aggregation, allows to reduce the communication load.

We therefore arrange the PEs in a logical two-dimensional grid as shown in
Fig.~\ref{fig:indirect}, and call the PE located in row $i$ and column $j$
$P_{i, j}$. When this PE wants to send a message to $P_{k, l}$, it first sends
it along the processor row to the so called \emph{proxy PE}, $P_{i, l}$. The
proxy then forwards the message to $P_{k, l}$ along the processor column.

To give an intuition on how this improves message flow, consider the extreme
case that all PEs want to send a message of size $1$ to a single destination PE.
Without message indirection, this PE has to receive $p$ messages, therefore
requiring time $p(\alpha + \beta)$. If indirection is employed, we double the
overall communication volume, but each PE only has $\sqrt{p}$ peers. This
results in a overall communication time of $\BigO(\sqrt{p}(\alpha + \beta))
+ p\beta$. This especially comes into play for large values of $p$.

Since each PE maintains a message queue, all messages from a processor row designated to $P_{k, l}$ get aggregated at the proxy. Using a threshold on the message queue as described before, we can still guarantee that each PE only requires $\BigO(|E_i|)$ space for aggregating messages.

If the number of PEs $p$ is not a square number, we arrange the PEs in a rectangular grid with $\lfloor\sqrt{p} + \frac{1}{2}\rfloor$ columns (i.e., we round to the nearest integer). 
The last row may not be completely filled. Suppose that $P_{i^\prime, j^\prime}$ wants to send a message to $P_{k, l}$ as depicted in Fig.~\ref{fig:indirect}.
Then the logical proxy does not exist.
We therefore transpose the last row and append it as a column to the right side of the grid
and then choose the proxy along the row as described before. 
Note that this is only necessary when sending from $P_{i^\prime, j^\prime}$ to $P_{k, l}$ and not in the other direction.

While this is somewhat similar to the indirect communication approach used in ~\cite{Pearce2019}, our grid-based redirection scheme is network-topology agnostic.

\subsection{Exploiting Locality}
\label{sec:algorithm}

We now propose a variant of \textsc{Ditric} called \textsc{Cetric}, a contraction based two-phase algorithm which counts triangles locally without using communication whenever possible.
The total communication volume of our algorithm is proportional to the size of the contracted graph consisting only of cut edges.

\begin{figure}
  \includegraphics[page=1, width=\columnwidth]{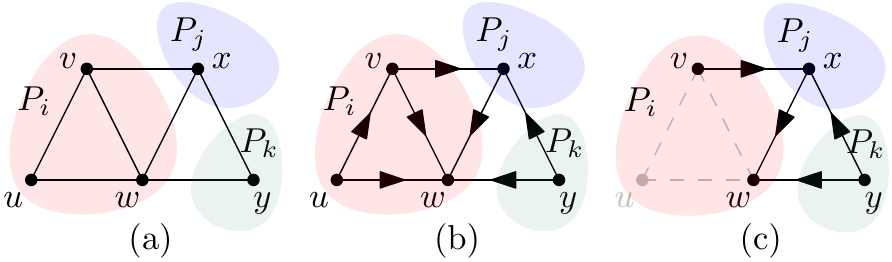}
  \caption{The main idea behind \textsc{Cetric}. (a) shows the different triangle types, $\{u, v, w\}$ is a type 1 triangle, $\{v, w, x\}$ a type 2 triangle and $\{w, x, y\}$ a type 3 triangle. (b) The edges are oriented towards high degree vertices. (c) After the local phase, all non cut edges can be removed from the graph.}
  \label{fig:cetric}
\end{figure}

We observe that each triangle $\{u, v, w\}$ in the input graph falls into one of the following categories: If all vertices are local to a single processor $P_i$, , i.e., if $u, v, w \in V_i$, we call it a \emph{type 1 triangle}. If any vertex $u \in V_j$, $j \neq i$ and both other vertices $v, w \in V_i$, we call it a \emph{type 2 triangle}. If each vertex in the triangle is local to a distinct processor, we call it a \emph{type 3 triangle}. The different triangle types are depicted in Fig.~\ref{fig:cetric}a).

Note that any type 1 triangle may always be found without communication, as all edges of the triangle are \enquote{known} to a single PE.
While this is also true for type 2 triangles, the basic distributed edge iterator algorithm introduced in section~\ref{sec:distributed-memory} does not leverage this due to the orientation of the edges. Consider the type 2 triangle $\{v, w, x\}$ in Fig.~\ref{fig:cetric}b). When PE $P_i$ examines edges $(v, w)$ the triangle is not found, because the algorithm only intersects outgoing neighborhoods. The triangle is only found by $P_j$ after $N_v^+$ has been sent by $P_i$. While this could be fixed by including the in- and outgoing neighborhoods of interface vertices in the set intersection, this would undo the effects of degree orientation, which reduces the out-degree of high-degree vertices.

We propose a two-phase algorithm which finds all type 1 and type 2 triangles using only locally available information while preserving degree orientation. We give a high level description of this approach in Algorithm~\ref{alg:basic}.

Our algorithm leverages this observation by using two phases.
In the \emph{local phase} our algorithm works on the \emph{expanded local graph} which consists of the set vertex set $\overline{V_i} = V_i \cup \partial V_i$, i.e., all local vertices and ghosts, and all edges which have at least one endpoint in $V_i$. Note that constructing this graph requires no communication, as it only consists of edges incident to local vertices. It is easy to see that running any sequential triangle counting algorithm on the expanded graphs yields all type 1 and type 2 triangles.

After the local phase we apply a \emph{contraction step}, which removes all non-cut edges from the graph as depicted in Fig.~\ref{fig:cetric}c).

In the \emph{global phase}, we can then use \textsc{Ditric} or any other distributed algorithm on the contracted cut graph to count the remaining type 3 triangles.

\begin{figure}
\removelatexerror
\begin{algorithm}[H]
	\DontPrintSemicolon
	\SetKwData{Sum}{SUM}
	\SetKwFunction{Red}{Reduce}
	\SetKw{Send}{send}
	\SetKw{To}{to}
	\SetKw{Recv}{receive}
	\SetKwFunction{AllToAll}{SparseAllToAll}
	\SetKwFor{OnReceive}{on receive}{do}{end}
	\SetKwProg{Fn}{function}{}{end}
	\SetKwFunction{CountTriangles}{count\_triangles}
	
    \tcp{Preprocessing (see sec.~\ref{sec:preprocessing})}
		\ExDeg{}\;
		$T_i \leftarrow 0$\;
		\lForEach{$v \in V_i$}{$A(v) \leftarrow \{x \in N_v \mid x \succ v\}$}
		\lForEach{$v \in \partial V_i$}{$A(v) \leftarrow \{x \in N_v \mid x \succ v \land x \in V_i \}$}
		\tcp{Local Phase}
		\ForEach{$v \in V_i \cup \partial V_i $}{
			\ForEach{$u \in A(v)$}{
					$T_i \leftarrow T_i + |A(v) \cap A(u)|$\; \label{line:intersection-cetric}
			}
			
		}
		\tcp{Contraction}
		\lForEach{$v \in V_i$}{$A(v) \leftarrow \{x \in N_v \mid x \succ v\} \setminus V_i $}
		\tcp{Global Phase}
		\ForEach{$v \in V_i$ with $A(v) \neq \emptyset$}{	
			\ForEach{$u \in A(v)$}{
				$j \leftarrow \rank(u)$\;
				\If{$A(v)$ not sent to $P_j$ yet}{ \label{line:surrogate}
					\Send $(v, A(v))$ \To $P_j$\;
				}
			}
			
		}
		\OnReceive{$(v, A(v))$}{
			\For{$u \in A(v)$ such that $u \in V_i$}{ 
				$T_i \leftarrow T_i + |A(v) \cap A(u)|$\;
			}
		}
		$T \leftarrow \Red{$T_i$, \Sum}$
	\caption{High-level overview \textsc{Cetric}. $i$ is the rank of the PE.}
	\label{alg:basic}
\end{algorithm}
\end{figure}

The following Lemma shows that removing all non-cut edges ensures that the global phase only finds type 3 triangles and that our algorithm is correct.

\begin{Lemma}
	\label{lemma:global-correctness}
	The vertex set $\{u, v, w\} \subseteq V$ induces a triangle in the cut graph $\partial G$ if and only if it is a type 3 triangle in $G$.
\end{Lemma}
\begin{IEEEproof}
	\begin{itemize}
		\item[$\Rightarrow$] Assume that $\{u, v, w\}$ induces a triangle in $G$
      that is not a type 3 triangle. Then there exist at least two endpoints of
      the triangle which belong to the same processor. W.l.o.g assume that these
      vertices are $u$ and $v$. Therefore, the edge $\{u, v\}$ connects two local vertices and is not contained in $\partial G$. This implies that at least one edge of the triangle from $G$ is missing in $\partial G$ and the triangle will not be counted.
		\item[$\Leftarrow$] Let $\{u, v, w\}$ be a type 3 triangle in $G$. Then each
      edge of the triangle connects vertices located on different processors.
      Therefore, each edge of the triangle is a cut edge and also contained in $\partial G$.
	\end{itemize}
\end{IEEEproof}

Due to the contraction, the communication volume of \textsc{Cetric} is only
dependent on the structure of the cut graph, while previous algorithms always
send the complete neighborhood of vertices to other PEs.

This can be combined with hybrid parallelism (i.e., using thread parallelism at
the compute node level) to achieve higher locality. We discuss this further in
section~\ref{sec:hybrid}.

\subsection{Implementation Details}
\label{sec:implementation-details}
In this section we discuss additional aspects of our implementation.

\paragraph*{Preprocessing}
\label{sec:preprocessing}
The preprocessing phase of our algorithms is responsible for applying the 
degree-based orientation and sorting vertex neighborhoods.
Orienting the edges from low to high-degree vertices requires each PE to
retrieve the degree of its ghost vertices. This is denoted by the procedure in
\texttt{exchange\_ghost\_degree} in Algorithm~\ref{alg:basic}.
This requires an all-to-all message exchange. If the number of communication partners is relatively low, this can benefit from using a sparse all-to-all operation~\cite{Hoefler2009}, which only sends direct messages to all its communication partners in a non-blocking way while continuously polling for incoming messages.
Preliminary experiments have shown that if the input has a skewed degree distribution, this may perform worse than a dense degree exchange. While one could use the indirect communication protocol proposed in section~\ref{sec:indirect}, we use a simple dense all-to-all operation in our evaluation, because the performance gains from using indirect communication for the initial degree exchanges are often small in the grand scheme.

Once the ghost degrees have been exchanged, we can easily construct the degree oriented graph. For \textsc{Cetric} we also need to expand the adjacency array structure for storing the local neighborhoods of ghost vertices. This requires no additional memory, because it simply means rewiring incoming cut edges to their corresponding ghost vertex.

\paragraph*{Avoiding redundant messages}
As stated before, extra care has to be taken to avoid sending the neighborhood of a vertex to the same PE multiple times.
We use the \emph{surrogate approach} introduced by Arrifuzzaman et al.~\cite{Arifuzzaman2015}, which relies on the global sortedness of vertex IDs and neighborhoods.
For each local vertex they keep track of the last PE the vertex's neighborhood has been sent to. 
When examining a cut edge $(v, u)$ in line~\ref{line:surrogate} in Algorithm~\ref{alg:basic} we only check if the last PE $A(v)$ has been sent to is not equal to $\rank{u}$. 
If this is the case, we encountered a new neighbor PE and enqueue the message for sending.

\paragraph*{Load Balancing}
Arifuzzaman et al.~\cite{Arifuzzaman2015} performed an extensive evaluation of load balancing for distributed triangle counting. They evaluate several degree-based cost functions which estimate the amount of work required to process each node and use a prefix-sum based redistribution of vertices among the PEs. When a new vertex distribution has been computed, they reload the graph from disk and do not account for this reloading step in the overall running time. We conducted preliminary experiments using their approach and adapted it to redistribute the graph using message passing, but observed that the overhead of rebalancing does not pay off.

\paragraph*{Hybrid parallelism}
\label{sec:hybrid}
We are currently working on extending our algorithms to a hybrid approach which uses multiple threads per MPI rank. 
Recall that our distributed-memory algorithm uses a 1D partitioning of the vertices of the input graph.
If a graph has a skewed degree distribution and therefore high-degree vertices, this may lead to work imbalances.
To reduce the effects of this, we use an adaptive approach on the thread level.
Instead of partitioning the local subgraph based on vertices, we partition the edge list consisting of local edges during the local phase. 
For each directed edge $(u, v)$ we perform the intersection in Line~\ref{line:intersection-cetric} in Algorithm~\ref{alg:basic} in parallel.

Previously, Green et al.~\cite{Green2014} reported good load balancing using this edge-centric parallelization strategy, but also noted that the preprocessing for distributing work evenly is slower than for a vertex-centric approach. By using work stealing, this can be omitted.

We were already able to achieve good speedups by parallelizing the local phase
of \textsc{Cetric} and notice a communication reduction by up to factor of 6
with 48 threads due to improved locality.

The hybridization of the global phase is realized using task stealing with
Intel's Thread Building Blocks library~\cite{TBB}.
Due to the limited scalability of current MPI implementations in full
multi-threaded environments, we restrict communication to one thread at a time per
MPI rank (using MPI's \emph{funneled} mode).
A pool of worker threads is responsible for handling tasks which write outgoing messages to the message
buffer. An additional thread continuously polls for incoming messages and creates
set intersections tasks from them which are pushed to the task queue of the workers. They prioritize
these tasks over writing additional data.

Preliminary experiments indicate that the communication thread becomes a
bottleneck, making the hybrid implementation slower than the plain MPI variant,
even though the local phase is faster and the communication volume is lower when
using the same number of compute nodes but using fewer MPI rank with
multi-threading enabled.
The results are show in the~\hyperref[app:hybrid-eval]{Appendix}.

\subsection{Extensions}
\label{sec:extensions}
So far we have only explained global triangle counting.
Since each triangle is found exactly once, this can be easily generalized to the case of
triangle enumeration.

We next explain how to determine the number of triangles $\Delta(v)$ incident to each vertex $v$.
In turn, this allows us to compute the local clustering coefficient of $v$ as $\mathrm{LCC}(v)=\frac{\Delta(v)}{d_v(d_v-1)}$.
Our algorithms find a triangle $\{v,u,w\}$ when iterating from exactly one incident vertex.
Then $\Delta(v)$, $\Delta(u)$, and $\Delta(w)$ have to be incremented. This is trivial for local triangles (type 1).
To make this also work for the distributed case, we also define $\Delta(v)$ for ghost vertices where it counts the triangles incident to $v$ that were found on that PE. As a postprocessing phase, we need to aggregate the $\Delta$-values of a vertex, including all its ghosts.
This can be implemented with an all-to-all exchange analogous to the initial degree exchange described in Section~\ref{sec:implementation-details}.

\textsc{Cetric} can also be made even more communication efficient at the cost
of computing only an approximation of the triangle count as follows:
Type~1~and~2 triangles are counted exactly. For type~3 triangles,
rather than sending a neighborhood $A(v)$ of a node, we only approximate it as $A'(v)$ using
an approximate membership query data structure (AMQ)%
\footnote{A typical implementation would be a Bloom Filter. We mention however, that a compressed single shot Bloom filter \cite{Putze2010} might be a more appropriate implementation here since it requires less communication volume.} A set intersection $A(u)\cap A(v)$ is then approximated by querying
all members of $A(u)$ in the AMQ $A'(v)$ counting the positive queries. Since AMQs have a certain likelihood of yielding a false-positive result, this will slightly overestimate the number of triangles.
If desired, we can correct for this by subtracting the expectation of the number of false-positives
from the count, yielding a truthful estimator.
This new approach to approximate triangle counting is particularly interesting when we want to approximate local clustering coefficients as the (faster) methods mentioned in Section~\ref{subsec:approx} are only applicable to global triangle counting.

\section{Evaluation}
\label{sec:evaluation}
\subsection{Experimental Setup}
We implemented our algorithm variants in C++. Our implementation is available at
\url{https://github.com/niklas-uhl/katric}.
We conduct our experiments using the thin nodes of 
the SuperMUC-NG supercomputer at the Leibniz Supercomputing Center.
The thin nodes consist of \num{} islands and a total of \num{6336} nodes, resulting in a total of \num{304128} cores.
Each node is equipped with an Intel Skylake Xeon Platinum 8174 processor with \num{48} cores.
The available memory per node is limited to \SI{96}{\giga\byte}. Each node runs the SUSE Linux Enterprise Server (SLES) operating system.
The internal interconnect is a fast OmniPath network with \SI{100}{\giga\bit\per\second}.

Our code is compiled with g++-11.2.0 and Intel MPI 2021 using optimization level \texttt{-O3}. 

\subsection{Methodology}
We analyze the scalability of our algorithm in terms of strong and weak scaling. For \emph{strong scaling} experiments we choose an input instance with fixed size and measure the running time over all processors with increasing number of processors. \emph{Weak scaling} measures the variation of running time for a fixed problem size \emph{per PE}, i.e., we scale the problem size proportional to the number of processors.

We evaluate several variants of our algorithms: 
\textsc{Ditric} and \textsc{Cetric} denote the simple distributed algorithm
using our dynamic message aggregation technique, without and with contraction,
respectively. The variants \textsc{Ditric$^2$} and \textsc{Cetric$^2$}
additionally use indirect messaging.

We compare our algorithm against the two latest champions of the Static Graph
Challenge~\cite{Samsi2017}. HavoqGT~\cite{Pearce2019} uses a vertex-centric
approach. They use a topology dependent routing protocol: To reduce the number
of messages, they first aggregate messages at node level and then reroute them
to other compute nodes. They also employ neighborhood partitioning of high
degree vertices among multiple PEs. TriC~\cite{Ghosh2020} uses static message
aggregation and does not orient the input graph.

If not stated otherwise, we report the total running time of each algorithm
excluding the time for reading the input graph and building the graph data
structure. For our algorithms, we still include the time for orienting edges and
sorting neighborhoods. Due to limitations of the used graph generators, we only
use core numbers which are a power of 2. For HavoqGT we only use (and count) 32
cores per compute node instead of 48, because it requires an equal amount of MPI
ranks per node. Since restricting all our experiments to 32 cores per node would
have resulted in many unused computing resources, we only conducted preliminary
experiments restricting our algorithms to 32 cores per node. They indicate that
this does not affect scalability in the grand scheme.

\subsection{Datasets}
We consider a variety of real-world and synthetic instances from different graph families which are among the largest publicly available.
All instances are listed in Table~\ref{tab:instances}.

\begin{table}[ht]
	\caption{Real-world graph instances used in our experiments.}
	\label{tab:instances}
	\centering
	\begin{tabular}{
			c
			l 
			S[table-format=3.0, round-mode=places, round-precision=0,  fixed-exponent=6, table-omit-exponent]
			@{\,}
			>{\collectcell\si}r<{\endcollectcell}
			S[table-format=4.0, round-mode=places, round-precision=0, fixed-exponent=6, table-omit-exponent]
			@{\,}
			>{\collectcell\si}l<{\endcollectcell}
			S[table-format=6.0, round-mode=places, round-precision=0, fixed-exponent=6, table-omit-exponent]
			@{\,}
			>{\collectcell\si}l<{\endcollectcell}
      S[table-format=6.0, round-mode=places, round-precision=0, fixed-exponent=6, table-omit-exponent]
      @{\,}
      >{\collectcell\si}l<{\endcollectcell}
		}
		& instance     & \multicolumn{2}{r}{$n$} & \multicolumn{2}{r}{$m$} & \multicolumn{2}{r}{wedges}   & \multicolumn{2}{r}{triangles}   \\ \midrule
		\multirow{4}{*}{\rotatebox[origin=c]{90}{social}} & live-journal & 4847571   & \mega\nothing  & 42851237   & \mega\nothing & 681198612 & \mega\nothing  & 285730264 & \mega\nothing \\
		& orkut        & 3072441   & \mega\nothing  & 117185083  & \mega\nothing & 4039818961 & \mega\nothing & 627584181 & \mega\nothing \\
		& twitter      & 41652230  & \mega\nothing  & 1202513046 & \mega\nothing & 150507617453 & \mega\nothing & 34824916864 & \mega\nothing \\
		& friendster   & 68349466  & \mega\nothing  & 1811849342 & \mega\nothing & 82285678111 & \mega\nothing & 4176922719 & \mega\nothing \\ \midrule[.1pt]
		\multirow{2}{*}{\rotatebox[origin=c]{90}{web}}    & uk-2007-05   & 105896555 & \mega\nothing  & 3301876564 & \mega\nothing & 389061279654 & \mega\nothing & 286701284103 & \mega\nothing \\
		& webbase-2001 & 118142155 & \mega\nothing  & 854809761  & \mega\nothing & 15393335873 & \mega\nothing & 12262060053 & \mega\nothing \\ \midrule[.1pt]
		\multirow{2}{*}{\rotatebox[origin=c]{90}{road}}   & europe       & 18029721  & \mega\nothing  & 22217686   & \mega\nothing & 7618963 & \mega\nothing & \multicolumn{2}{r}{\num{697519}} \\
		& usa          & 23947347  & \mega\nothing  & 28854312   & \mega\nothing & 10767843 & \mega\nothing & \multicolumn{2}{r}{\num{438804}}
	\end{tabular}
	
\end{table}

We use live-journal and orkut from the SNAP dataset~\cite{Leskovec2014}, a complete snapshot of follower relationships of the twitter network~\cite{Kwak2010}\footnote{available for download at \url{http://an.kaist.ac.kr/traces/WWW2010.html}} and friendster from the KONECT collection~\cite{Kunegis2013}\footnote{available for download at \url{http://konect.cc}}. We further include large web graphs from the Laboratory for Web Algorithmics dataset collection (LAW), namely uk-2007-05 and webbase-2001~\cite{Boldi2004, Boldi2011}\footnote{available for download at \url{http://law.di.unimi.it/datasets.php}}. Graph uk-2007-05 has a size of 50 GB and has over 3 billion edges and is the largest real-world graph used in our experiments. It is the result of a crawl of the \texttt{.uk} domain. webbase-2001 has been originally obtained from the 2001 crawl performed by the WebBase crawler.  While having more vertices than uk-2007-05, it only has around 855 million edges.
We use road networks of Europe and USA made available during the DIMACS challenge on shortest paths~\cite{Demetrescu2009}. Road networks typically have low average degree and a uniform degree distribution.
Whenever the graphs are directed, we interpret each edge as undirected. We remove vertices with no neighbors from the input.

For our weak scaling experiments we generate synthetic graph instances
using \textsc{KaGen} \cite{Funke2019,HuebschleSchneider2020}.
This allows us to generate very large graphs in a controllable way and without the need to load them from the file system which is quite expensive on a supercomputer.
Particularly, we use 2D random geometric graphs (RGG2D), random hyperbolic graphs (RHG), Erdős–Rényi graphs using the $G(n, m)$ model (GNM) and R-MAT graphs.

For all random graph models we generate \num{16} times more edges than vertices (on expectation), which is the default used in the Graph 500 benchmark~\cite{Graph}.
The 2D random geometric graph model places $n$ vertices at uniform random locations in the unit square. Two vertices are adjacent if their euclidean distance is less than a given radius $r$. We choose $r$ such that the expected number of edges is $16n$.
RHG graphs are generated by placing vertices on a disk with fixed radius $R$, which is dependent on the desired average degree and the power-law exponent $\gamma$, which also affects the concentration of points around the center of the disk. For our experiments we choose a power-law exponent of $\gamma = 2.8$.
In the $G(n, m)$ model, graphs are chosen uniformly at random from the set of all possible graphs with $n$ vertices and $m$ edges. 
The recursive matrix graph model (R-MAT) is used in the popular Graph 500 benchmark~\cite{Graph}. It recursively subdivides the adjacency array into four equally sized sectors with associated probabilities and recursively descends into one of the sectors. We use the default probabilities from the Graph 500 benchmark. 

\subsection{Weak Scaling Experiments}
\begin{figure*}
	\centering
		\input{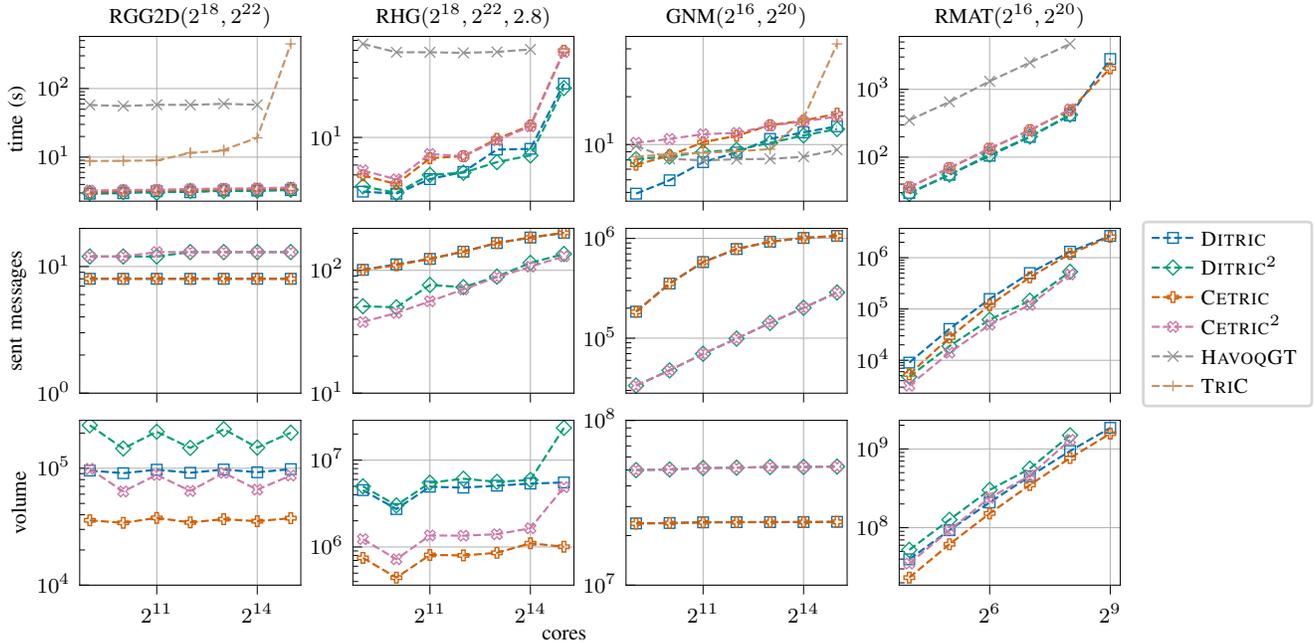}
	\caption{Weak scaling of our algorithms compared to the competitors using up
    to $2^{15}$ cores. We report total running time, the maximum number of outgoing
    messages over all PEs and bottleneck communication volume.}
	\label{fig:weak-scaling}
\end{figure*}

In Fig.~\ref{fig:weak-scaling} we show the results of our weak scaling experiments.
We report the total running time of all algorithms without the time for reading
the input. We also report the maximum number of outgoing messages over all PEs
and the bottleneck communication volume for our algorithms.
For RGG2D we set the input size per PE to $\frac{n}{p} = 2^{18}$ vertices.
We see that all our algorithms clearly outperform TriC and HavoqGT and that they all show similar scaling behavior. TriC fails to scale beyond $2^{14}$ cores. We only report results for HavoqGT up to $2^{14}$ cores, because its preprocessing time exceeded \SI{900}{\second}.
TriC only works on RGG2D and GNM. For the other instances it crashes due to high memory consumption. Perhaps this happens because it allocates the complete message buffer upfront and performs a single irregular all-to-all operation. Since the communication volume may be superlinear in the input size, this may exhaust a single PE's memory.

For random hyperbolic graphs, we choose the same input size as before.
We again outperform HavoqGT by an order of magnitude. We also see that both variants of \textsc{Ditric} slightly outperform \textsc{Cetric}, but still show the same scaling behavior. This comes as a surprise, because usually one expects communication volume to be the bottleneck of a distributed memory computation on large complex networks. 
Further investigation on real-world instances in the following section indicates that this is not always the case on supercomputers with high-performance network interconnects.
While our contraction based algorithm \textsc{Cetric} reduces the bottleneck communication volume by up to a factor of 4, it also requires a constant factor more local work, to also process cut edges. On these large graph instances, the local work clearly dominates the overall running time. We still expect \textsc{Cetric} to outperform \textsc{Ditric} on communication networks with higher latency and lower bandwidth.
From $2^{12}$ cores onward, we also see that the indirect communication approach
employed \textsc{Ditric}$^2$ is makes it on average \SI{10}{\percent} faster
than plain \textsc{Ditric}. The spike in running time for $2^{15}$ PEs is caused
by an increased time for the degree exchange as a result of the skewed degree distribution.

For GNM we choose $\frac{n}{p} = 2^{16}$ vertices per core.
Up to $2^{11}$ PEs \textsc{Ditric} is the fastest algorithm.
We see that the variants of \textsc{Cetric} are up to \SI{50}{\percent} slower than \textsc{Ditric}. This comes at no surprise, because random graphs have no locality. We therefore achieve almost no reduction of communication volume, but require additional local work, which does not pay off.
From $2^{11}$ cores onward, HavoqGT needs up to \SI{30}{\percent} less time than our algorithms.

Since the skewed degree distribution leads to
high running times, we conducted experiments for RMAT graphs only using smaller scale inputs and
lower processor configurations.
Our algorithms are an order of magnitude faster than HavoqGT, but all algorithms show worse scaling behavior on RMAT than on the other synthetic graph families. 
Again, the additional local work for contraction does not pay off.

\subsection{Strong Scaling Experiments}
\begin{figure}[t]
	\centering
	\centering
		\input{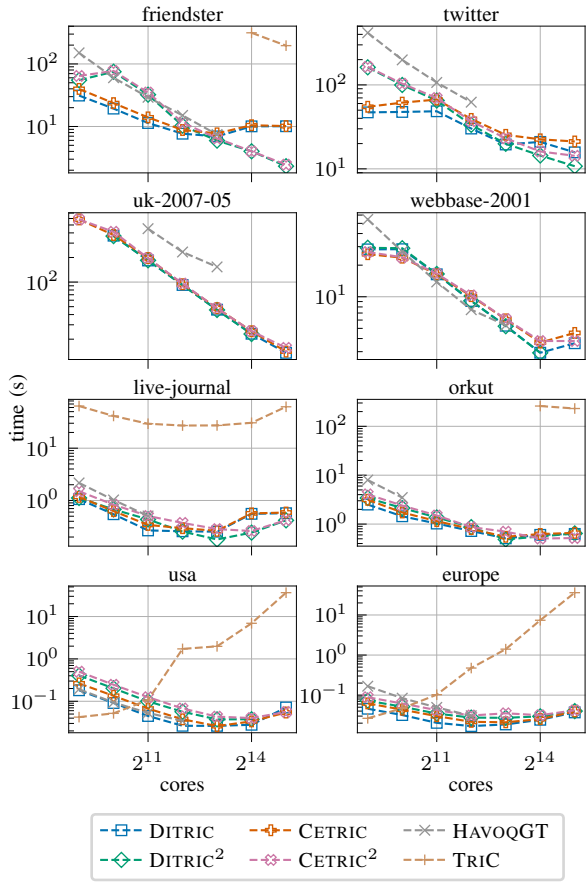}
	\caption{Strong scaling on real world instances using $2^9$ to $2^{15}$ cores.}
	\label{fig:strong-scaling}
\end{figure}
\begin{figure}[t]
	\centering
		\input{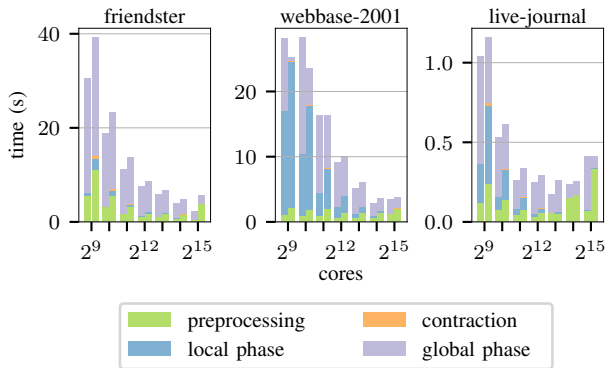}
	\caption{Running time distribution of our algorithms on selected real world instances. The left bar reports the running times for the best variant of \textsc{Ditric}, the right bar for the best variant of \textsc{Cetric}.}
	\label{fig:phase-times}
  \vspace{-.5cm}
\end{figure}

We present the results of the strong scaling experiments in Fig.~\ref{fig:strong-scaling} on a variety of real world graph instances, and Fig.~\ref{fig:phase-times} shows a detailed break-down of the algorithm phases for selected instances. We allowed to run each algorithm for 300 seconds on each instance, including I/O and preprocessing. Note that we integrated our I/O routines into the competitors for better comparison.

On the social networks of friendster and twitter, we see that \textsc{Ditric} is up to 8 times faster than HavoqGT. 
While our algorithm variants with indirect communication are slower than those using direct messages up to $2^{13}$ PEs, they show better scalability for large configurations. The effects of indirect communication become especially visible on friendster. Note that we only were able to run TriC using $2^{14}$ and $2^{15}$ PEs on friendster, where it is 80 times slower than our best algorithm. On other configurations we failed to execute TriC, because it ran out of memory. As stated before, we attribute this to the static buffering.
Examining Fig.~\ref{fig:phase-times} we see that on friendster, \textsc{Cetric} requires additional preprocessing time to build the expanded graph, but does not achieve to reduce communication by a large order of magnitude. We assume that this is due to the missing locality in the input graphs, which is exploited by \textsc{Cetric}.

On live-journal we are up to two orders of magnitude faster than TriC, and 2 times faster than HavoqGT with our fastest configuration. For more than $2^{11}$ PEs HavoqGT's preprocessing took more than 300 seconds, which is why we do not report results. From $2^{12}$ PEs onward, indirect communication improves the scalability.
Taking a closer look at the differences between \textsc{Ditric} and
\textsc{Cetric} in Fig.~\ref{fig:phase-times}, we see that \textsc{Cetric}
halves the time required in the global phase due to the reduced communication
volume. Unfortunately, to achieve this, it also requires additional preprocessing and local work, which ultimately does not pay off. We still expect our contraction-based algorithm variant to outperform \textsc{Ditric} on a system with slower networks interconnects. This may for example be the case in large cloud computing environments. 
We observe similar results for orkut.

On webbase-2001 we see that the variants of \textsc{Cetric} are faster than
\textsc{Ditric} up to $2^{11}$ PEs. In Fig.~\ref{fig:phase-times} wee observe that
while requiring additional local work, the communication reduction by almost a
factor of 2 pays off, but only up to a certain number of processors. With
increasing number of PEs the cut of the graphs grows bigger, allowing the removal of fewer edges during contraction. From $2^{12}$ PEs onward, we see that almost no reduction of the global phase is visible.

On road networks TriC is initially faster than our variants.
The graphs have small average degree and cut size, which leads to a small
communication volume. This profits from TriC's single batch communication, but fails to scale beyond
$2^{11}$ cores. From there on \textsc{Ditric} slightly
outperforms HavoqGT. This experiment may seem like using a sledgehammer to crack
a nut, because counting triangles on europe takes already less than a second
using 4 cores, but it shows that our algorithms do not hit a scaling wall, even on small inputs.

\section{Conclusion and Future Work}
\label{sec:conclusion}

We have engineered triangle counting codes that scale to many thousands of cores and are sufficiently communication efficient that the local computations dominate when using a high performance network. We achieved this by employing message aggregation and reducing the communication volume by exploiting locality.

We think it makes sense to use algorithms with superlinear communication volume such as triangle counting as additional standard benchmarks for high performance graph processing.

Both locality and scalability could be further enhanced by
improving the shared-memory part of the code such that 
it scales to all threads available in a compute node. It would also be interesting to develop a low-overhead load balancing algorithm that allows provable performance guarantees.
Further performance could be gained by using GPU-acceleration for local computations.
On the software side, it now seems important to achieve a similar level of performance in graph-processing tools that make it easier for non-HPC-experts to implement a variety of graph analysis tasks.

\section*{Acknowledgment}
The authors gratefully acknowledge the Gauss Centre for Supercomputing e.V. (\url{www.gauss-centre.eu}) for funding this project by providing computing time on the GCS Supercomputer SuperMUC-NG at Leibniz Supercomputing Centre (\url{www.lrz.de}).

\begin{figure}[H]
	\centering
  \input{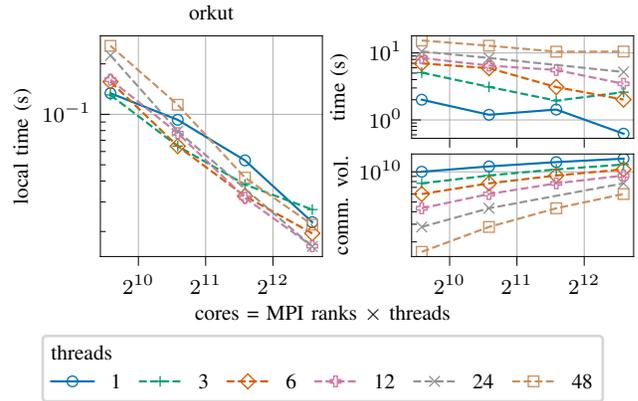}
	\caption{Local phase time, total time and communication volume for
    the hybrid variant of \textsc{Ditric}$^2$ using up to \num{6144} cores.}
	\label{fig:hybrid}
\end{figure}

\appendix[Evaluation of Hybrid Parallelism]
\label{app:hybrid-eval}
In Fig.~\ref{fig:hybrid} we show a preliminary evaluation of the hybridization
of our approach described in Section~\ref{sec:hybrid}. We report results for
\textsc{Ditric} with message indirection on orkut. We fix the number of physical
cores used, but vary the number of threads, such that $\text{cores} =
\text{threads} \times \text{MPI ranks}$.

We achieve a speedup of up to $1.67$ during the local phase with 12 threads over the single threaded
variant using the same number of PEs and reduce the communication volume by up to $84\%$, but the
hybridization of the global phase of our algorithm becomes a bottleneck which
limits the overall scalability.

\IEEEtriggeratref{28}

\bibliographystyle{IEEEtran}
\bibliography{IEEEabrv,references}

\begin{thebibliography}{10}
\providecommand{\url}[1]{#1}
\csname url@samestyle\endcsname
\providecommand{\newblock}{\relax}
\providecommand{\bibinfo}[2]{#2}
\providecommand{\BIBentrySTDinterwordspacing}{\spaceskip=0pt\relax}
\providecommand{\BIBentryALTinterwordstretchfactor}{4}
\providecommand{\BIBentryALTinterwordspacing}{\spaceskip=\fontdimen2\font plus
\BIBentryALTinterwordstretchfactor\fontdimen3\font minus
  \fontdimen4\font\relax}
\providecommand{\BIBforeignlanguage}[2]{{%
\expandafter\ifx\csname l@#1\endcsname\relax
\typeout{** WARNING: IEEEtran.bst: No hyphenation pattern has been}%
\typeout{** loaded for the language `#1'. Using the pattern for}%
\typeout{** the default language instead.}%
\else
\language=\csname l@#1\endcsname
\fi
#2}}
\providecommand{\BIBdecl}{\relax}
\BIBdecl

\bibitem{Sanders2023}
P.~Sanders and T.~N. Uhl, ``Engineering a distributed-memory triangle counting
  algorithm,'' in \emph{2023 IEEE International Parallel and Distributed
  Processing Symposium (IPDPS)}, 2023, pp. 702--712.

\bibitem{Statista2021}
Statista, ``Number of monthly active facebook users worldwide as of 4th quarter
  2020,''
  \url{https://www.statista.com/statistics/264810/number-of-monthly-active-facebook-users-worldwide/},
  Jan. 2021.

\bibitem{Meusel}
R.~Meusel, O.~Lehmberg, C.~Bizer, and S.~Vigna, ``Web data commons - hyperlink
  graph,''
  \url{http://km.aifb.kit.edu/sites/webdatacommons/hyperlinkgraph/index.html}.

\bibitem{Graph}
{The Graph 500 steering commitee}, ``The {Graph 500} benchmark,''
  \url{https://graph500.org}.

\bibitem{Samsi2017}
S.~Samsi, V.~Gadepally, M.~Hurley, M.~Jones, E.~Kao, S.~Mohindra,
  P.~Monticciolo, A.~Reuther, S.~Smith, W.~Song, D.~Staheli, and J.~Kepner,
  ``Static {Graph} {Challenge}: {Subgraph} {Isomorphism},'' \emph{IEEE High
  Performance Extreme Computing {Conf.} (HPEC)}, pp. 1--6, 2017.

\bibitem{Newman2003}
M.~E.~J. Newman, ``The structure and function of complex networks,'' \emph{SIAM
  review}, vol.~45, no.~2, pp. 167--256, 2003.

\bibitem{Becchetti2008}
L.~Becchetti, P.~Boldi, C.~Castillo, and A.~Gionis, ``Efficient semi-streaming
  algorithms for local triangle counting in massive graphs,'' in \emph{19th ACM
  {Conf.} on Knowledge Discovery and Data Mining}, 2008, pp. 16--24.

\bibitem{Eckmann2002}
J.-P. Eckmann and E.~Moses, ``Curvature of co-links uncovers hidden thematic
  layers in the world wide web,'' \emph{Proc. of the National Academy of
  Sciences}, vol.~99, no.~9, pp. 5825--5829, 2002.

\bibitem{BarYossef2002}
Z.~Bar-Yossef, R.~Kumar, and D.~Sivakumar, ``Reductions in streaming
  algorithms, with an application to counting triangles in graphs,'' in
  \emph{Symposium on Discrete Algorithms (SODA)}, vol.~2, 2002, pp. 623--632.

\bibitem{Tsourakakis2011}
C.~E. Tsourakakis, P.~Drineas, E.~Michelakis, I.~Koutis, and C.~Faloutsos,
  ``Spectral counting of triangles via element-wise sparsification and
  triangle-based link recommendation,'' \emph{Social Network Analysis and
  Mining}, vol.~1, no.~2, pp. 75--81, 2011.

\bibitem{Schank2007}
T.~Schank, ``Algorithmic aspects of triangle-based network analysis,'' Ph.D.
  dissertation, {University Karlsruhe (TH)}, 2007.

\bibitem{Ortmann2014}
M.~Ortmann and U.~Brandes, ``Triangle listing algorithms: Back from the
  diversion,'' in \emph{SIAM Symposium on Algorithm Engineering and Experiments
  ({ALENEX})}, 2014, pp. 1--8.

\bibitem{Latapy2008}
M.~Latapy, ``Main-memory triangle computations for very large (sparse
  (power-law)) graphs,'' \emph{Theoretical computer science}, vol. 407, no.
  1-3, pp. 458--473, 2008.

\bibitem{Shun2015}
J.~Shun and K.~Tangwongsan, ``Multicore triangle computations without tuning,''
  in \emph{IEEE 31st {Conf.} on Data Engineering}, 2015, pp. 149--160.

\bibitem{Dhulipala2021}
L.~Dhulipala, G.~E. Blelloch, and J.~Shun, ``Theoretically {Efficient}
  {Parallel} {Graph} {Algorithms} {Can} {Be} {Fast} and {Scalable},'' \emph{ACM
  Transactions on Parallel Computing}, vol.~8, pp. 4:1--4:70, 2021.

\bibitem{Tom2017}
A.~S. Tom, N.~Sundaram, N.~K. Ahmed, S.~Smith, S.~Eyerman, M.~Kodiyath, I.~Hur,
  F.~Petrini, and G.~Karypis, ``Exploring optimizations on shared-memory
  platforms for parallel triangle counting algorithms,'' in \emph{IEEE High
  Performance Extreme Computing Conf. (HPEC)}, 2017, pp. 1--7.

\bibitem{Rahman2013}
M.~Rahman and M.~Al~Hasan, ``Approximate triangle counting algorithms on
  multi-cores,'' in \emph{{IEEE} Conf. on Big Data}, 2013, pp. 127--133.

\bibitem{Green2014}
O.~Green, L.-M. Mungu{\'\i}a, and D.~A. Bader, ``Load balanced clustering
  coefficients,'' in \emph{Proc. 1st workshop on Parallel Programming for
  Analytics Applications}, 2014, pp. 3--10.

\bibitem{Arifuzzaman2015}
S.~Arifuzzaman, M.~Khan, and M.~Marathe, ``A space-efficient parallel algorithm
  for counting exact triangles in massive networks,'' in \emph{IEEE 17th
  {Intl.} {Conf.} on High Performance Computing and Communications}, 2015, pp.
  527--534.

\bibitem{Ghosh2020}
S.~Ghosh and M.~Halappanavar, ``{TriC}: Distributed-memory triangle counting by
  exploiting the graph structure,'' in \emph{{IEEE} High Performance Extreme
  Computing {Conf.} ({HPEC})}, 2020, pp. 1--6.

\bibitem{Tom2019}
A.~S. Tom and G.~Karypis, ``A {2D} parallel triangle counting algorithm for
  distributed-memory architectures,'' in \emph{48th ACM Conf. on Parallel
  Processing}, 2019.

\bibitem{Azad2015}
A.~Azad, A.~Bulu{\c c}, and J.~Gilbert, ``Parallel triangle counting and
  enumeration using matrix algebra,'' in \emph{IEEE Parallel and Distributed
  Processing Symposium Workshop}, 2015, pp. 804--811.

\bibitem{Suri2011}
S.~Suri and S.~Vassilvitskii, ``Counting triangles and the curse of the last
  reducer,'' in \emph{20th {Conf.} on World Wide Web}, 2011, pp. 607--614.

\bibitem{Dean2004}
J.~Dean and S.~Ghemawat, ``{MapReduce}: Simplified data processing on large
  clusters,'' in \emph{6th Symposium on Operating Systems Design and
  Implementation}, 2004.

\bibitem{Park2014}
H.-M. Park, F.~Silvestri, U.~Kang, and R.~Pagh, ``Mapreduce triangle
  enumeration with guarantees,'' in \emph{23rd ACM {Conf.} on Information and
  Knowledge Management}, 2014, pp. 1739--1748.

\bibitem{Afrati2012}
F.~N. Afrati, A.~D. Sarma, S.~Salihoglu, and J.~D. Ullman, ``Upper and lower
  bounds on the cost of a map-reduce computation,'' \emph{arXiv preprint
  arXiv:1206.4377}, 2012.

\bibitem{Pearce2014}
R.~Pearce, M.~Gokhale, and N.~M. Amato, ``Faster parallel traversal of scale
  free graphs at extreme scale with vertex delegates,'' in \emph{IEEE {Conf.}
  for High Performance Computing, Networking, Storage and Analysis}, 2014, pp.
  549--559.

\bibitem{Pearce2017}
R.~Pearce, ``Triangle counting for scale-free graphs at scale in distributed
  memory,'' in \emph{IEEE High Performance Extreme Computing {Conf.} (HPEC)},
  2017, pp. 1--4.

\bibitem{Pearce2019}
R.~Pearce, T.~Steil, B.~W. Priest, and G.~Sanders, ``One {Quadrillion}
  {Triangles} {Queried} on {One} {Million} {Processors},'' in \emph{{IEEE}
  {High} {Performance} {Extreme} {Computing} {Conf.} ({HPEC})}, 2019, pp. 1--5.

\bibitem{Hoang2019}
L.~Hoang, V.~Jatala, X.~Chen, U.~Agarwal, R.~Dathathri, G.~Gill, and
  K.~Pingali, ``{DistTC}: {High} {Performance} {Distributed} {Triangle}
  {Counting},'' in \emph{{IEEE} {High} {Performance} {Extreme} {Computing}
  {Conf.} ({HPEC})}, 2019, pp. 1--7.

\bibitem{Arifuzzaman2013}
S.~Arifuzzaman, M.~Khan, and M.~Marathe, ``{PATRIC}: A parallel algorithm for
  counting triangles in massive networks,'' in \emph{Proceedings of the 22nd
  ACM international conference on Information \& Knowledge Management}, 2013,
  pp. 529--538.

\bibitem{Tsourakakis2009}
C.~E. Tsourakakis, U.~Kang, G.~L. Miller, and C.~Faloutsos, ``Doulion: counting
  triangles in massive graphs with a coin,'' in \emph{15th ACM {Conf.} on
  Knowledge Discovery and Data Mining}, 2009, pp. 837--846.

\bibitem{Pagh2012}
R.~Pagh and C.~E. Tsourakakis, ``Colorful triangle counting and a mapreduce
  implementation,'' \emph{Information Processing Letters}, vol. 112, no.~7, pp.
  277--281, 2012.

\bibitem{Jha2013}
M.~Jha, C.~Seshadhri, and A.~Pinar, ``A space efficient streaming algorithm for
  triangle counting using the birthday paradox,'' in \emph{19th ACM {Conf.} on
  Knowledge Discovery and Data Mining}, 2013, pp. 589--597.

\bibitem{AlHasan2018}
M.~Al~Hasan and V.~S. Dave, ``Triangle counting in large networks: a review,''
  \emph{WIREs Data Mining and Knowledge Discovery}, vol.~8, no.~2, 2018.

\bibitem{Green2014a}
O.~Green, P.~Yalamanchili, and L.-M. Mungu{\'\i}a, ``Fast triangle counting on
  the {GPU},'' in \emph{Proc. 4th Workshop on Irregular Applications:
  Architectures and Algorithms}, 2014, pp. 1--8.

\bibitem{Hu2018}
Y.~Hu, H.~Liu, and H.~H. Huang, ``{TriCore}: {Parallel} {Triangle} {Counting}
  on {GPUs},'' in \emph{{Intl.} {Conf.} for {High} {Performance} {Computing},
  {Networking}, {Storage} and {Analysis}}, 2018, pp. 171--182.

\bibitem{Pandey2019}
S.~Pandey, X.~S. Li, A.~Buluc, J.~Xu, and H.~Liu, ``H-{INDEX}:
  {Hash}-{Indexing} for {Parallel} {Triangle} {Counting} on {GPUs},'' in
  \emph{{IEEE} {High} {Performance} {Extreme} {Computing} {Conf.} ({HPEC})},
  2019, pp. 1--7.

\bibitem{Chu2011}
S.~Chu and J.~Cheng, ``Triangle listing in massive networks and its
  applications,'' in \emph{17th ACM {Conf.} on Knowledge Discovery and Data
  Mining}, 2011, pp. 672--680.

\bibitem{Hoefler2009}
T.~Hoefler and J.~L. Traff, ``Sparse collective operations for {MPI},'' in
  \emph{{IEEE} {Intl.} {Symposium} on {Parallel} {Distributed} {Processing}},
  2009, pp. 1--8.

\bibitem{TBB}
\BIBentryALTinterwordspacing
``{I}ntel one{API} {T}hreading {B}uilding {B}locks.'' [Online]. Available:
  \url{https://github.com/oneapi-src/oneTBB}
\BIBentrySTDinterwordspacing

\bibitem{Putze2010}
F.~Putze, P.~Sanders, and J.~Singler, ``Cache-, hash-, and space-efficient
  bloom filters,'' \emph{{ACM} Journal of Experimental Algorithmics}, vol.~14,
  2009.

\bibitem{Leskovec2014}
J.~Leskovec and A.~Krevl, ``{SNAP Datasets}: {Stanford} large network dataset
  collection,'' \url{http://snap.stanford.edu/data}, 2014.

\bibitem{Kwak2010}
H.~Kwak, C.~Lee, H.~Park, and S.~Moon, ``{W}hat is {T}witter, a social network
  or a news media?'' in \emph{19th {Conf.} on World Wide Web}, 2010, pp.
  591--600.

\bibitem{Kunegis2013}
J.~Kunegis, ``{KONECT} -- {The} {Koblenz} {Network} {Collection},'' in
  \emph{Proc. Intl. Conf. on World Wide Web Companion}, 2013, pp. 1343--1350.

\bibitem{Boldi2004}
P.~Boldi and S.~Vigna, ``The webgraph framework {I:} compression techniques,''
  in \emph{Intl. World Wide Web Conference ({WWW})}, 2004, pp. 595--602.

\bibitem{Boldi2011}
P.~Boldi, M.~Rosa, M.~Santini, and S.~Vigna, ``Layered label propagation: A
  multiresolution coordinate-free ordering for compressing social networks,''
  in \emph{20th {Conf.} on World Wide Web}, 2011, pp. 587--596.

\bibitem{Demetrescu2009}
C.~Demetrescu, A.~V. Goldberg, and D.~S. Johnson, \emph{The Shortest Path
  Problem: Ninth DIMACS Implementation Challenge}, ser. Dimacs Series in
  Discrete Mathematics and Theoretical Computer Science, 2009, vol.~74.

\bibitem{Funke2019}
D.~Funke, S.~Lamm, U.~Meyer, M.~Penschuck, P.~Sanders, C.~Schulz, D.~Strash,
  and M.~von Looz, ``Communication-free massively distributed graph
  generation,'' \emph{Journal of Parallel and Distributed Computing}, vol. 131,
  pp. 200--217, 2019.

\bibitem{HuebschleSchneider2020}
L.~H{\"u}bschle-Schneider and P.~Sanders, ``Linear work generation of {R-MAT}
  graphs,'' \emph{Network Science}, vol.~8, no.~4, pp. 543--550, 2020.

\end{thebibliography}

\end{document}